\documentclass[11pt,a4,twocolumn,secnumarabic,amssymb, amsmath, nofootinbib,tightenlines,
nobibnotes, aps, prl,epsfig]{revtex4}
\usepackage{graphicx}
\usepackage{dcolumn}
\usepackage{bm}
\begin{document}
\preprint{APS/123-QED}
\title{Decoupling of the DGLAP evolution equations at next-to-next-to-leading order (NNLO) at low-$x$ }
\author{G.R.Boroun}
 \email{grboroun@gmail.com; boroun@razi.ac.ir}
\author{B.Rezaei }
\altaffiliation{brezaei@razi.ac.ir}
\affiliation{ Physics Department, Razi University, Kermanshah
67149, Iran}
\date{\today}

\begin{abstract}
We present a set of formulae to extract two second-order
independent differential equations for the gluon and  singlet
distribution functions. Our results extend from the LO up to NNLO
DGLAP evolution equations with respect to the hard- Pomeron
behavior at low-$x$. In this approach, both singlet quarks and
gluons have the same high-energy behavior at low-$x$. We solve the
independent DGLAP evolution equations for the functions
$F_{2}^{s}(x,Q^{2})$ and $G(x,Q^{2})$  as a function of their
initial parameterisation at the starting scale $Q_{0}^{2}$. The
results not only give striking support to the hard- Pomeron
description of the low-$x$ behavior, but give a rather clean test
of perturbative QCD showing an increase of the gluon distribution
and singlet structure functions as $x$ decreases. We compared our
numerical results with the published BDM (M.M.Block, L.Durand and
D.W.Mckay, Phys.Rev.D77, 094003(2008)) gluon and singlet
distributions, starting from their initial values at $Q_{0}^{2}=1
\mathrm{GeV}^{2}$.
\end{abstract}
\maketitle
\section{Introduction}
The Dokshitzer-Gribov-Lipatov-Altarelli-Parisi (DGLAP) [1]
evolution equations are fundamental tools to study the $Q^{2}$-
and $x$-evolutions of structure functions, where $x$ is the
Bjorken scaling parameter and $Q ^{2}$ is the virtuality of the
exchanged vector boson in a deep inelastic scattering process [2].
The measurements of the $F_{2}(x,Q ^{2} )$ structure functions by
deep inelastic scattering processes in the low-$x$ region have
opened a new era in parton density measurements inside hadrons.
The structure function reflects the momentum distributions of the
partons in the nucleon. It is also important to know the gluon
distribution inside a hadron at low-$x$ because gluons are
expected to be dominant in this region. The steep increase of
$F_{2}(x,Q ^{2} )$ towards low-$x$ observed at the hadron electron
ring accelerator (HERA) also indicates a similar increase in the
gluon distribution towards low-$x$ in perturbative quantum
chromodynamics. In the usual procedure, the deep inelastic
scattering data are analyzed by the next-to-next-to-leading order
(NNLO) QCD fits based on the numerical solution of the DGLAP
evolution equations, and it has been found that the DGLAP analysis
can well describe the data in the perturbative region $Q ^{2} \geq
1 \mathrm{GeV} ^{2}$ [3]. As an alternative to the numerical
solution, one can study the behavior of quarks and gluons via
analytic solutions of the evolution equations. Although exact
analytic solutions of the DGLAP equations cannot be obtained in
the entire range of $x$ and $Q ^{2}$, such solutions are possible
under certain conditions and are quite successful as far as the
HERA low-$x$ data are concerned. Some of these  methods [4] were
proposed in the
literature by using expansion method or pomeron behavior.\\
The low-$x$ region of DIS offers a unique possibility to explore
the Regge limit of pQCD [5]. This theory is successfully described
by the exchange of a particle with appropriate quantum numbers and
the exchanged particle is called a Regge pole. Phenomenologically,
the Regge pole approach to DIS implies that the structure
functions are sums of powers in $x$, modulus logarithmic terms,
each with a $Q^{2}$- dependent residue factor. Also, in the DGLAP
formalism the gluon splitting functions are singular as $x
\rightarrow 0$. Thus, the gluon distribution will become large as
$x \rightarrow 0$, and its contribution to the evolution of the
parton distribution becomes dominant. In particular, the gluon
will drive the quark singlet distribution, and, hence, the
structure function $F_{2}$ becomes large as well, the rise
increasing in steepness as $Q^{2}$ increases. This model gives the
following parametrization of the DIS parton distribution functions
$xf_{k}(x,Q^{2}) (k=\Sigma,g)$ at low-$x$ where $f_{k}(x,Q^{2})$
is the parton density. This phenomenon is usually described by
assuming a power-like behavior of parton distribution functions as
$xf_{k}(x,Q^{2}) = f_{k}(Q^{2})x^{-\delta} $, that the singlet
part of the parton distribution functions are controlled by
Pomeron exchange at low-$x$, where $\delta$ is the Pomeron
intercept minus one. For $Q^{2} \leq 1 \mathrm{GeV}^{2}$, the
simplest Regge phenomenology predicts that the value of $\delta =
\alpha_{\mathbb{P}}(0)- 1 \simeq 0.08 $ is consistent with that of
hadronic Regge theory, where $\alpha_{\mathbb{P}}(0)$ is described
by soft- Pomeron dominant with its intercept slightly above unity
( ${\sim} 1.08$), whereas for $Q^{2}\geq 1 \mathrm{GeV}^{2}$ the
slope rises steadily to reach a value greater than 0.4 by $Q^{2}
\approx 100 \mathrm{GeV}^{2}$,
where hard-Pomeron is dominant [5-7].\\

The one-loop splitting functions corresponding to LO DGLAP
equation are given in Ref.[8]. Similarly the two-loop splitting
functions governing the evolution have been known for a long term
[9]. The effects of NLO [9] and NNLO [10-14] terms in the
evolution parton structure functions are known to be important,
especially at low-$x$ in the gluon and singlet sector. The
calculation of the NNLO QCD approximation for the parton structure
functions of DIS is important for the understanding of
perturbative QCD (PQCD) and for an accurate comparison of PQCD
with experiment. To obtain the NNLO approximation for these parton
structure functions one needs the corresponding three-loop
splitting functions. Traditionally, gluon and quark distribution
functions have been determined using the two coupled
integral-differential (DGLAP) equations to evolve individual quark
and gluon distributions. Here, we propose a new method for
determining the gluon and quark distribution functions by using
the two decoupled homogeneous second-order differential equation
which determine individual $G(x,Q^{2})$ and $F_{2}(x,Q^{2})$,
respectively. In the evolution parton structure functions and
running coupling we take $N_{f}=4$ for $m_{c}<\mu<m_{b}$, which at
the starting scale of evolution at $Q_{0}^{2}$, we use the Block
fit [15,16] to ZEUS data [17] in the domain $10^{-3} \leq x \leq
0.09 $ and $0.11 \leq Q^{2} \leq 1200
\mathrm{GeV}^{2}$.\\

The analytical methods of the unpolarized DGLAP evolution
equations have been discussed considerably in $x$-space, Mellin
and Laplace transformation [18,19,15]. Some approximated
analytical solutions of DGLAP evolution equations suitable at
low-$x$, have been reported in last years [4] with considerable
phenomenological success. The distributions have been obtained
using the coupled DGLAP evolution equations, in LO and NLO.
Recently, in Ref.[20] decoupled solutions of the LO and NLO
coupled DGLAP evolution equations have been obtained using Laplace
transformation. Those results show that obtained solutions deepen
on both initial condition of the gluon distribution function and
singlet structure function at the initial scale. The decoupled
solutions of the NLO DGLAP evolution equations (with respect to
the Taylor series expanding and the  hard-Pomeron behavior) found
in Ref.[21] at low-$x$, where the gluon kernel is dominant. In the
present paper, such solutions can be generalised to NNLO by
solving the decoupled DGLAP evolution equations at low-$x$ as both
gluon and singlet kernels are dominant. In this paper, we will
study the decoupling DGLAP evolution equations based on the
hard-Pomeron behavior of the gluon and individual quark
distributions. The method gives a global gluon and quark
distribution function in the $x$ and $Q^{2}$ space which depend
explicitly on the gluon and quark distribution individual at
$Q_{0}^{2}$ scale, respectively.\\
\subsection{Theory}
The HERA data should determine the low-$x$ behavior of gluon and
singlet quark distributions. We will be concerned specifically
with the singlet contribution to the proton structure function at
LO, as
\begin{eqnarray}
F^{ep}_{2}(x,Q^{2}){\equiv}x\sum_{i=1}^{N_{f}}e_{i}^{2}(q_{i}(x,Q^{2})+\overline{q}_{i}(x,Q^{2})),
\end{eqnarray}
where $N_{f}$ is the number of active flavors. At low-$x$ and
high-$Q^{2}$ the singlet quark distribution is essentially driven
by the generic instability of the gluon distribution
$G(x,Q^{2})=xg(x,Q^{2})$, where $g(x,Q^{2})$ is the gluon density.
To see how this works, consider the singlet Altarelli- Parisi
equations [1], which describe perturbative evolution of
$xg(x,Q^{2})$ and
$x\Sigma(x,Q^{2})$.\\
The DGLAP evolution equations for the singlet quark structure
function and the gluon distribution are given by
\begin{widetext}
\begin{equation}
\frac{{\partial}G(x,Q^{2})}{{\partial}{\ln}Q^{2}}=\frac{\alpha_{s}}{2\pi}{\int_{x}^{1}}dz[
P_{gg}(z,\alpha_{s}(Q^{2}))
G(\frac{x}{z},Q^{2})+P_{gq}(z,\alpha_{s}(Q^{2}))
\Sigma(\frac{x}{z},Q^{2})]
\end{equation}
\begin{equation}
\frac{{\partial}\Sigma(x,Q^{2})}{{\partial}{\ln}Q^{2}}=\frac{\alpha_{s}}{2\pi}{\int_{x}^{1}}dz[
P_{qq}(z,\alpha_{s}(Q^{2}))
\Sigma(\frac{x}{z},Q^{2})+2N_{f}P_{qg}(z,\alpha_{s}(Q^{2}))
G(\frac{x}{z},Q^{2})]
\end{equation}
\end{widetext}
where  $\Sigma(x,Q^{2})$ and $G(x,Q^{2})$ are singlet and gluon
distribution functions, and the splitting functions $P_{ij}^{,}s$
are the LO, NLO and NNLO Altarelli- Parisi splitting kernels as
\begin{widetext}
\begin{eqnarray}
P_{ij}(x,\alpha_{s}(Q^{2}))=P_{ij}^{\rm
LO}(x)+\frac{\alpha_{s}(Q^{2})}{2\pi}P_{ij}^{\rm NLO}(x)
+(\frac{\alpha_{s}(Q^{2})}{2\pi})^{2} P_{ij}^{\rm NNLO}(x).
\end{eqnarray}
\end{widetext}
The next-to-leading order is the standard approximation for most
important processes. The corresponding one- and two-loop splitting
functions have been known for a long time. The NNLO corrections
need to be included, however, in order to obtain a quantitatively
reliable predictions for hard processes at present and future
high-energy colliders. These corrections are so far known  for
structure functions in deep-inelastic scattering (DIS) [22] and
for Drell-Yan lepton-pair [23].\\
The quark-quark splitting function $P_{qq}$ in Eq.~(3) can be
expressed as
$P_{qq}=P^{+}_{ns}+N_{f}(P_{qq}^{s}+P_{\overline{q}q}^{s}){\equiv}P^{+}_{ns}+P_{ps}$.
Here $P^{+}_{ns}$ is the non-singlet splitting function which at
low-$x$ is negligible and can be ignored . $P^{s} _{qq}$ and
$P^{s} _{\overline{q}q}$ are the flavour independent contributions
to the quark-quark and quark-antiquark splitting functions,
respectively. At low-$x$, the pure singlet term $P_{ps}$ dominates
over $P^{+}_{ns}$ [12-13]. The gluon-quark ($P_{gq}$) and
quark-gluon ($P_{qg}$) entries in Eqs.~(2) and (3) are given by
$P_{qg}=N_{f}P_{q_{i}g}$ and $P_{gq}=P_{gq_{i}}$ where
$P_{q_{i}g}$ and $P_{gq_{i}}$ are the flavor-independent splitting
functions.\\
The running coupling constant $\frac{\alpha_{s}}{2\pi}$ has the
form in the LO, NLO and NNLO respectively [24]
\begin{equation}
\frac{\alpha_{s}^{\rm LO}}{2\pi}=\frac{2}{\beta_{0}t},
\end{equation}
\begin{equation}
\frac{\alpha_{s}^{\rm
NLO}}{2\pi}=\frac{2}{\beta_{0}t}[1-\frac{\beta_{1}{\ln}t}{\beta_{0}^{2}t}],
\end{equation}
and
\begin{eqnarray}
\frac{\alpha_{s}^{\rm
NNLO}}{2\pi}&=&\frac{2}{\beta_{0}t}[1-\frac{\beta_{1}{\ln}t}{\beta_{0}^{2}t}+\frac{1}{(\beta_{0}t)^{2}}
[(\frac{\beta_{1}}{\beta_{0}})^{2}\nonumber\\
&&(\ln^{2}t-{\ln}t+1)+\frac{\beta_{2}}{\beta_{0}}]].
\end{eqnarray}
where $\beta_{0}=\frac{1}{3}(33-2N_{f})$,
$\beta_{1}=102-\frac{38}{3}N_{f}$ and
$\beta_{2}=\frac{2857}{6}-\frac{6673}{18}N_{f}+\frac{325}{54}N_{f}^{2}$
are the one-loop,two-loop and three-loop corrections to the QCD
$\beta$-function. The variable $t$ is defined as
$t={\ln}(\frac{Q^{2}}{\Lambda^{2}})$ and $\Lambda$ is the QCD
cut- off parameter.\\
\subsection{Decoupling solutions at LO}
The  LO DGLAP evolution equations for the gluon distribution
function and the proton structure function  for massless quarks
can be written as
\begin{widetext}
\begin{equation}
\frac{{\partial}G(x,Q^{2})}{{\partial}{\ln}Q^{2}}=\frac{\alpha^{\rm
LO}_{s}}{2\pi}{\int_{x}^{1}}dz P^{\rm LO}_{gg}(z)
G(\frac{x}{z},Q^{2})+\frac{18}{5}\frac{\alpha^{\rm
LO}_{s}}{2\pi}{\int_{x}^{1}}dzP^{\rm LO}_{gq}(z)
F_{2}(\frac{x}{z},Q^{2}),
\end{equation}
\begin{equation}
\frac{{\partial}F_{2}(x,Q^{2})}{{\partial}{\ln}Q^{2}}=\frac{\alpha^{\rm
LO}_{s}}{2\pi}{\int_{x}^{1}}dz P^{\rm LO}_{qq}(z)
F_{2}(\frac{x}{z},Q^{2})+\frac{20}{9}\frac{\alpha^{\rm
LO}_{s}}{2\pi}{\int_{x}^{1}}dzP^{\rm LO}_{qg}(z)
G(\frac{x}{z},Q^{2}).
\end{equation}
\end{widetext}
Since
$F^{ep}_{2}(x,Q^{2})=\frac{5}{18}\Sigma(x,Q^{2})+\frac{3}{18}F^{NS}_{2}(x,Q^{2})$,
we should be able to ignore the non-singlet contribution
$F^{NS}_{2}(x,Q^{2})$ to the proton structure function at low-$x$
values. Now let us introduce the hard-Pomeron behavior for the
parton structure functions. As it is well known, the parton
structure functions obtained from fits to data follow an
approximate power-law behavior [6-7] at low-$x$,
\begin{equation}
F_{2}(x,t)=f_{F}(t)x^{-\delta}, G(x,t)=f_{G}(t)x^{-\delta}
\end{equation}
at given $t$, where $f_{i}$ depend of course on the parton species
and $\delta$ is taken as a hard trajectory intercept mines one.
The power $\delta$ is found to be either $\delta{\simeq}0$ or
$\delta{\simeq}0.5$ [25]. The first value corresponds to the soft
Pomeron and the second value the hard (Lipatov) Pomeron intercept.\\
Using (10) in (8) and (9) and performing z-integrations, we have
\begin{widetext}
\begin{equation}
\frac{{\partial}G(x,t)}{{\partial}t}=\frac{\alpha^{\rm
LO}_{s}}{2\pi}G(x,t){\int_{x}^{1}}dz P^{\rm LO}_{gg}(z)z^{\delta}
+\frac{18}{5}\frac{\alpha^{\rm
LO}_{s}}{2\pi}F_{2}(x,t){\int_{x}^{1}}dzP^{\rm
LO}_{gq}(z)z^{\delta},
\end{equation}
\begin{equation}
\frac{{\partial}F_{2}(x,t)}{{\partial}t}=\frac{\alpha^{\rm
LO}_{s}}{2\pi}F_{2}(x,t){\int_{x}^{1}}dz P^{\rm
LO}_{qq}(z)z^{\delta} +\frac{20}{9}\frac{\alpha^{\rm
LO}_{s}}{2\pi}G(x,t){\int_{x}^{1}}dzP^{\rm LO}_{qg}(z)z^{\delta}.
\end{equation}
\end{widetext}
After some rearranging, we find two homogeneous second-order
differential equations which determine $F_{2}(x,t)$ and $G(x,t)$
without having knowledge in terms of $G(x,t)$ and $F_{2}(x,t)$,
respectively. As we have
\begin{eqnarray}
\frac{\partial^{2}{{\digamma}(x,t)}}{\partial{t}^{2}}+\frac{1}{t}\eta(x)\frac{\partial{{\digamma}(x,t)}}{\partial{t}}
+\frac{1}{t^{2}}\zeta(x)\digamma(x,t)=0,\nonumber\\
(\digamma=F_{2}\hspace{0.1cm} or \hspace{0.1cm}G)\hspace{3cm}
\end{eqnarray}
To simplify the notation in Eq.~13, we define the initial
conditions by
\begin{eqnarray}
\digamma_{0}{\equiv}\digamma(x,t_{0}),
\digamma{\digamma}_{0}=\frac{\partial{{\digamma}(x,t)}}{\partial{t}}|_{t=t_{0}}.
\end{eqnarray}
Therefore, the analytic solution for the proton structure function
and gluon distribution function with respect to the initial
conditions and those derivatives can be obtained as
\begin{widetext}
\begin{eqnarray}
{\digamma}(x,t)=\frac{1}{2}\frac{(\eta(x)\digamma_{0}-\digamma_{0}+\sqrt{\eta(x)^{2}-2\eta(x)+1-4\zeta(x)}\digamma_{0}+2\digamma{\digamma}_{0}t_{0})}
{\sqrt{\eta(x)^{2}-2\eta(x)+1-4\zeta(x)}}
(\frac{t}{t_{0}})^{\frac{1}{2}(-\eta(x)+1+\sqrt{\eta(x)^{2}-2\eta(x)+1-4\zeta(x)})}\nonumber\\
+\frac{1}{2}\frac{(-\eta(x)\digamma_{0}+\digamma_{0}+\sqrt{\eta(x)^{2}-2\eta(x)+1-4\zeta(x)}\digamma_{0}-2\digamma{\digamma}_{0}t_{0})}
{\sqrt{\eta(x)^{2}-2\eta(x)+1-4\zeta(x)}}
(\frac{t}{t_{0}})^{\frac{1}{2}(-\eta(x)+1-\sqrt{\eta(x)^{2}-2\eta(x)+1-4\zeta(x)})}\nonumber\\
\end{eqnarray}
\end{widetext}
These results are completely general and gives the exact functions
of $x$ and $t$(or $Q^{2}$) in a domain
$x_{min}{\leq}x{\leq}x_{max}$ and
$Q^{2}_{min}{\leq}Q^{2}{\leq}Q^{2}_{max}$. The explicit forms of
the functions $\eta(x)$ and $\zeta(x)$ are given in Appendix A.\\
\subsection{Decoupling solutions at NLO up to NNLO}
With respect to the hard-Lipatove Pomeron behavior of the
structure function and gluon distribution function and
substituting the splitting functions up to NLO and up to NNLO in
DGLAP evolution equations we have
\begin{eqnarray}
\frac{{\partial}F_{2}(x,t)}{{\partial}t}+M(x,t) F_{2}(x,t)+N(x,t) G(x,t)=0,\nonumber\\
\frac{{\partial}G(x,t)}{{\partial}t}+P(x,t) G(x,t)+Q(x,t)
F_{2}(x,t)=0.\nonumber\\
\end{eqnarray}
where the explicit forms of the functions and up to third-order
splitting functions are given by in Appendix B. The decoupling
solutions of the coupled equations in Eqs.~(16) in terms of the
initial conditions are straightforward. After successive
differentiations of both equations of Eq.~(16) and some
rearranging, we find two homogeneous second-order differential
equation for the structure function and gluon distribution
function respectively,
\begin{widetext}
\begin{eqnarray}
\frac{{\partial}^{2}F_{2}(x,t)}{{\partial}t^{2}}+[N(x,t)\frac{{\partial}}{{\partial}t}(\frac{1}{N(x,t)})+M(x,t)+P(x,t)]\frac{{\partial}F_{2}(x,t)}{{\partial}t}\nonumber\\
+[N(x,t)\frac{{\partial}}{{\partial}t}(\frac{M(x,t)}{N(x,t)})+P(x,t)M(x,t)-P(x,t)Q(x,t)] F_{2}(x,t)=0,\nonumber\\
\frac{{\partial}^{2}G(x,t)}{{\partial}t^{2}}+[Q(x,t)\frac{{\partial}}{{\partial}t}(\frac{1}{Q(x,t)})+M(x,t)+P(x,t)]\frac{{\partial}G(x,t)}{{\partial}t}\nonumber\\
+[Q(x,t)\frac{{\partial}}{{\partial}t}(\frac{P(x,t)}{Q(x,t)})+P(x,t)M(x,t)-N(x,t)Q(x,t)]G(x,t)=0.
\end{eqnarray}
\end{widetext}
These results are completely general and give the exact NLO and
NNLO expression with respect to the running coupling constant
(Eqs.~(6) and (7)) and the splitting functions (Eq.~(4)) up to NLO
and up
to NNLO respectively.\\
\subsection{Results and Discussion}
In this paper, we found two analytical decoupled solutions for the
coupled DGLAP evolution equations for the proton structure
function and the gluon distribution function inside the proton.
These decoupling equations are directly related to the initial
conditions and to the strong interaction coupling constant at LO,
NLO and NNLO. To determine the proton structure function and gluon
distribution function we need to know only the input singlet and
gluon densities and their derivatives at the initial scale of
$Q_{0}^{2}$, respectively. The input singlet and gluon
parameterizations can be taken from global analysis of the parton
distribution functions, in particular from the Block analysis
[15,16]. We furthermore follow the DL model [6,7] in taking the
hard-Pomeron intercept with $\delta{\simeq}0.5$. We will compare
the $x$-space structure function and gluon distribution function
calculated from Eqs.15 and 17 at LO up to NNLO starting from the
Block initial conditions at $Q_{0}^{2}=1 \mathrm{GeV}^{2}$. We
will also compare our
results with H1 data [26] numerically.\\
In Figs.1 and 2, we show the results for the proton structure
function and gluon distribution function at LO up to NNLO at
$Q^{2}=20 \mathrm{GeV}^{2}$. The solid curve in these figures is
the published Block method [15-16] and also dots are H1 data [26]
that accompanied with total errors in Fig.1. In these figures, the
squares, down triangles and up triangles are our results for LO,
NLO and NNLO from Eqs.15 and 17. We present the results using the
$F_{0}(x)$ and $G_{0}(x)$ which are usually taken from the Block
model. The agreement between our results at NNLO analysis and the
Block method is good. It is clear from these figures for
$F_{2}(x,Q^{2})$ and $G(x,Q^{2})$, that our decoupling solutions
are correct. As can be seen, the values of the gluon distribution
and the proton structure functions increase as $x$ decreases, this
is because the hard-Pomeron exchange defined by the DL model is
expected to hold in the low-$x$ limit. It is evident from Figs.1
and 2 that three-loop perturbative QCD describes the evolution of
the strength of the hard-Pomeron contribution to $F_{2}(x,Q^{2})$
and $G(x,Q^{2})$ very well
 with respect to the decoupling DGLAP evolution equations.\\
\subsection{Conclusion}
We have first developed a method for the analytic solution of the
DGLAP evolution equations based on the hard-Pomeron behavior of
the parton distributions at low-$x$. In conclusion, we have
constructed two decoupled homogeneous second-order differential
evolution equations for $F_{2}(x,Q^{2})$ and $G(x,Q^{2})$ from the
coupled DGLAP equations at LO up to NNLO analysis, respectively.
These results for the gluon distribution and proton structure
functions  require only a knowledge individual from $G_{0}(x)$,
$F_{0}(x)$ and those derivatives at the starting value $Q_{0}^{2}$
for the evolution, respectively. As an illustration of our method,
we have used the analytic solutions to the decoupled evolution
equations to obtain tests of the consistency our results with
published quark and gluon distributions. We demonstrated
numerically that the method gives good agreement with published
Block method and H1 data at
NNLO.\\

$\bf{Acknowledgements}$   G.R.Boroun would like to thank the
anonymous referee of the paper for his/her careful reading of the
manuscript and for the productive discussions.\\

\subsection{Appendix A}
The explicit forms of the functions $a(x)$, $b(x)$, $c(x)$ and
$d(x)$ are
\begin{eqnarray}
&&\eta(x)=1-a(x)-c(x).\nonumber\\
&& \zeta(x)=a(x)c(x)-b(x)d(x).\nonumber\\
&&a(x)=\frac{2}{\beta_{0}}{\int_{x}^{1}}dzP^{LO}_{qq}(z)z^{\delta}.\nonumber\\
&&b(x)=\frac{2}{\beta_{0}}{\int_{x}^{1}}dzP^{LO}_{qg}(z)z^{\delta}.\nonumber\\
&&c(x)=\frac{2}{\beta_{0}}{\int_{x}^{1}}dzP^{LO}_{gg}(z)z^{\delta}.\nonumber\\
&&d(x)=\frac{2}{\beta_{0}}{\int_{x}^{1}}dzP^{LO}_{gq}(z)z^{\delta}.
\end{eqnarray}
Where the splitting functions are given by [5,27]
\begin{eqnarray}
&&P^{\rm LO}_{qq}(z)=C_{F}[\frac{1+z^{2}}{(1-z)_{+}}+\frac{3}{2}\delta(1-z)].\nonumber\\
&&P^{\rm LO}_{qg}(z)=\frac{1}{2}(z^{2}+(1-z)^{2}).\nonumber\\
&&P^{\rm LO}_{gq}(z)=C_{F}\frac{1+(1-z)^2}{z}.\nonumber\\
&&P^{\rm LO}_{gg}(z)=2C_{A}(\frac{z}{(1-z)_{+}}+\frac{(1-z)}{z}+z(1-z))\nonumber\\
&&+\delta(1-z)\frac{(11C_{A}-4N_{f}T_{R})}{6},
\end{eqnarray}
with $C_{A}=N_{c}=3$,
$C_{F}=\frac{N_{c}^{2}-1}{2N_{c}}=\frac{4}{3}$ and
 $T_{f}=\frac{1}{2}N_{f}$. The convolution integrals in (18) which
contains plus prescription, $()_{+}$, can be easily calculate by
[28]
\begin{eqnarray}
\int_{x}^{1}\frac{dy}{y}f(\frac{x}{y})_{+}g(y)&=&\int_{x}^{1}\frac{dy}{y}f(\frac{x}{y})[g(y)-\frac{x}{y}g(x)]\nonumber\\
&&-g(x)\int_{0}^{x}f(y)dy\nonumber\\
\end{eqnarray}

\subsection{Appendix B}
The explicit forms of the functions $M(x,t), N(x,t), P(x,t)$ and
$Q(x,t)$ are
\begin{eqnarray}
M(x,t)=-\frac{\alpha_{s}}{2\pi}\int_{x}^{1}{P_{qq}(z,\alpha_{s}(Q^{2}))z^{\delta}dz}.\nonumber\\
N(x,t)=-\frac{\alpha_{s}}{2\pi}\int_{x}^{1}{2N_{f}P_{qg}(z,\alpha_{s}(Q^{2}))z^{\delta}dz}.\nonumber\\
P(x,t)=-\frac{\alpha_{s}}{2\pi}\int_{x}^{1}{P_{gg}(z,\alpha_{s}(Q^{2}))z^{\delta}dz}.\nonumber\\
Q(x,t)=-\frac{\alpha_{s}}{2\pi}\int_{x}^{1}{P_{gq}(z,\alpha_{s}(Q^{2}))z^{\delta}dz}.
\end{eqnarray}
where the strong coupling constant,$\alpha_{s}$, up to NNLO is
given by Eqs.~(6-7). The explicit forms of the second- and third-
order splitting functions are respectively [12-14]
\begin{widetext}
\begin{eqnarray}
P_{qq}^{\rm NLO}&=&(C_F)^2(-1+z+(1/2-3/2z)\ln(z)-1/2(1+z)\ln(z)^{2}-(3/2\ln(z)+2\ln(z)\ln(1-z))p_{qq}(z)\nonumber\\
&&+2p_{qq}(-z)S_2(z))+C_F C_A(14/3(1-z)+(11/6\ln(z)+1/2\ln(z)^{2}+67/18-\pi^2/6)p_{qq}(z)\nonumber\\
&&-p_{qq}(-z)S_2(z))+C_FT_F(-16/3+40/3z+(10z+16/3z^2+2)\ln(z)-112/9z^2+40/(9z)\nonumber\\
&&-2(1+z)\ln(z)^{2}-(10/9+2/3\ln(z))p_{qq}(z)).\nonumber\\
P_{qg}^{\rm NLO}&=&C_FT_F(4-9z-(1-4z)\ln(z)-(1-2z)\ln(z)^{2}+4\ln(1-z)+(2\ln((1-z)/z)^{2}-4ln((1-z)/z)\nonumber\\
&&-2/3\pi^2+10)P_{qg}(z))+C_AT_F(182/9+14/9z+40/(9z)+(136/3z-38/3)\ln(z)-4\ln(1-z)\nonumber\\
&&-(2+8z)\ln(z)^{2}+2P_{qg}(-z)S_2(z)+(-\ln(z)^{2}+44/3\ln(z)-2\ln(1-z)^{2}+4\ln(1-z)\nonumber\\
&&+\pi^2/3-218/9)P_{qg}(z)).\nonumber\\
P_{gq}^{\rm NLO}&=&C_F^2(-5/2-7z/2+(2+7/2z)\ln(z)-(1-z/2)ln(z)^{2}-2z\ln(1-z)-(3\ln(1-z)\nonumber\\
&&+\ln(1-z)^{2})P_{gq}(z))+C_FC_A(28/9+65/18z+44/9z^2-(12+5z+8/3z^2)\ln(z)+(4+z)\ln(z)^{2}\nonumber\\
&&+2z\ln(1-z)+S_2(z)P_{gq}(-z)+(1/2-2\ln(z)\ln(1-z)+1/2\ln(z)^{2}+11/3\ln(1-z)+\ln(1-z)^{2}\nonumber\\
&&-\pi^2/6)P_{gq}(z))+C_FT_F(-4/3z-(20/9+4/3\ln(1-z))P_{gq}(z)).\nonumber\\
P_{gg}^{\rm NLO}&=&C_FT_F(-16+8z+20/3z^2+4/(3z)-(6+10z)\ln(z)-(2+2z)\ln(z)^{2})+C_AT_F(2-2z\nonumber\\
&&+26/9(z^2-1/z)-4/3(1+z)\ln(z)-20/9P_{gg}(z))+C_A^2(27/2(1-z)+26/9(z^2-1/z)\nonumber\\
&&-(25/3-11/3z+44/3z^2)\ln(z)+4(1+z)\ln(z)^{2}+2P_{gg}(-z)S_2(z)+(67/9-4\ln(z)\ln(1-z)\nonumber\\
&&+\ln(z)^{2}-\pi^2/3)P_{gg}(z)).\nonumber\\
\end{eqnarray}
\end{widetext}
where
\begin{eqnarray}
p_{qq}(z)=2/(1-z)-1-z\nonumber\\
p_{qq}(-z)=2/(1+z)-1+z\nonumber\\
P_{qg}(z)=z^2+(1-z)^2\nonumber\\
P_{qg}(-z)=z^2+(1+z)^2\nonumber\\
P_{gq}(z)=(1+(1-z)^2)/z\nonumber\\
P_{gq}(-z)=-(1+(1+z)^2)/z\nonumber\\
P_{gg}(z)=1/(1-z)+1/z-2+z(1-z)\nonumber\\
P_{gg}(-z)=1/(1+z)-1/z-2-z(1+z)\nonumber\\
S_2(z)=\int_{1/(1+z)}^{z/(1+z)}1/y\ln((1-y)/y)dy\nonumber\\
\end{eqnarray}
and
\begin{widetext}
\begin{eqnarray}
P_{qq}^{\rm NNLO}&=&(N_f(-5.926L1^3-9.751L1^2-72.11L1+177.4+392.9z-101.4z^2-57.04L0L1-661.6L0\nonumber\\
&&+131.4L0^2-400/9L0^3+160/27L0^4-506/z-3584/271/zL0)+N_f^2(1.778L1^2+5.944L1\nonumber\\
&&+100.1-125.2z+49.26z^2-12.59z^3-1.889L0L1+61.75L0+17.89L0^2+32/27L0^3\nonumber\\
&&+256/811/z))(1-z).\nonumber\\
P_{qg}^{\rm NNLO}&=&N_f(100/27L1^4-70/9L1^3-120.5L1^2+104.42L1+2522-3316z+2126z^2\nonumber\\
&&+L0L1(1823-25.22L0)-252.5zL0^3+424.9L0+881.5L0^2-44/3L0^3+536/27L0^4\nonumber\\
&&-1268.31/z-896/31/zL0)+N_f^2(20/27L1^3+200/27L1^2-5.496L1-252+158z+145.4z^2\nonumber\\
&&-139.28z^3-L0L1(53.09+80.616L0)-98.07zL0^2+11.70zL0^3-254L0-98.80L0^2-376/27L0^3\nonumber\\
&&-16/9L0^4+1112/2431/z).\nonumber\\
P_{gq}^{\rm NNLO}&=&400/81L1^4+2200/27L1^3+606.3L1^2+2193L1-4307+489.3z+1452z^2+146z^3-447.3L0^2L1\nonumber\\
&&-972.9zL0^2+4033L0-1794L0^2+1568/9L0^3-4288/81L0^4+6163.11/z+1189.31/zL0\nonumber\\
&&+N_f(-400/81L1^3-68.069L1^2-296.7L1-183.8+33.35z-277.9z^2+108.6zL0^2\nonumber\\
&&-49.68L0L1+174.8L0+20.39L0^2+704/81L0^3+128/27L0^4-46.411/z+71.0821/zL0)\nonumber\\
&&+N_f^2(96/27L1^2(1/z-1+1/2z)+320/27L1(1/z-1+4/5z)-64/27(1/z-1-2z)).\nonumber\\
P_{gg}^{\rm NNLO}&=&2643.524D0+4425.894\delta(1-z)+3589L1-20852+3968z-3363z^2+4848z^3\nonumber\\
&&+L0L1(7305+8757L0)+274.4L0-7471L0^2+72L0^3-144L0^4+142141/z+2675.81/zL0\nonumber\\
&&+N_f(-412.142D0-528.723\delta(1-z)-320L1-350.2+755.7z-713.8z^2+559.3z^3\nonumber\\
&&+L0L1(26.15-808.7L0)+1541L0+491.3L0^2+832/9L0^3+512/27L0^4+182.961/z\nonumber\\
&&+157.271/zL0)+N_f^2(-16/9D0+6.4630\delta(1-z)-13.878+153.4z-187.7z^2+52.75z^3\nonumber\\
&&-L0L1(115.6-85.25z+63.23L0)-3.422L0+9.680L0^2-32/27L0^3-680/2431/z).
\end{eqnarray}
\end{widetext}
where $L0=\ln(z), L1=\ln(1-z)$ and $D0=1/(1-z)$.\\

\newpage{
\subsection{References}
1. Yu. L.Dokshitzer, Sov.Phys.JETPG {\bf6}, 641(1977 );
G.Altarelli and
G.Parisi, Nucl.Phys.B{\bf126}, 298(1997 ); V.N.Gribov and L.N.Lipatov, Sov.J.Nucl.Phys.{\bf28}, 822(1978).\\
2. L.F.Abbott, W.B.Atwood and A.M.Barnett, Phys.Rev.D {\bf22}, 582(1980).\\
3. A.M.Cooper- Sarkar, R.C.E.Devenish and A.DeRoeck,
Int.J.Mod.Phys.A {\bf13}, 3385( 1998 ); M.Cluck, E.Reya and
A.Vogt, Z.Phys.C{\bf48},471 (1990); A.D.Martin, W.J.Stirling, and
R.S.Thorne, Phys.Lett.B{\bf636}, 259(2006); A.D.Martin,
W.J.Stirling, R.S.Thorne and G.Watt, Eur.Phys.J.C{\bf63},
189(2009); A.D.Martin, W.J.Stirling, R.S.Thorne and G.Watt,
arXiv:1301.6754(hep-ph); M. Gluck, P. Jimenez-Delgado, E. Reya,
Eur.Phys.J.C\textbf{53}, 355(2008); M. Gluck,
P. Jimenez-Delgado, E. Reya, Phys.Rev.D\textbf{79}, 074023(2009).\\
 4. K.Prytz, Phys.Lett.B{\bf311}, 286(1993); K.Prytz. Phys.Lett.B{\bf332}, 393(1994); M.B. Gay Ducati and V.P.B.Goncalves, Phys.Lett.B{\bf390}, 401(1997);
  A.V.Kotikov and G.Parente, Phys.Lett.B{\bf379}, 195(1996); P.Desgrolard, A.Lengyel and E.Martynov, JHEP {\bf02}, 029(2002); A.Donnachie and P.V.Landshoff,
   Phys.Lett.B \textbf{533}, 277(2002); J.R.Cudell, A.Donnachie and P.V.Landshoff, Phys.Lett.B \textbf{448}, 281(1999); J.Kwiecinski, arXiv:hep-ph/9607221(1996).\\
5. P.D.Collins, \textit{An introduction to Regge theory an
high-energy physics}(Cambridge University Press, Cambridge 1997)Cambridge; M.Bertini \textit{et al}., Rivista del Nuovo Cimento \textbf{19}, 1(1996).\\
6. A.Donnachie and P.V.Landshoff, Z.Phys.C \textbf{61}, 139(1994);
Phys.Lett.B \textbf{518}, 63(2001).\\
7.A.Donnachie and P.V.Landshoff, Phys.Lett.B \textbf{550},
160(2002); R.D.Ball and P.V.Landshoff, arXiv:hep-ph/9912445.\\
8. G.Altarelli, G.Parisi, Nucl.Phys.B \textbf{126}, 298(1977).\\
9. W.Furmanski, R.Petronzio, Phys.Lett.B \textbf{97}, 437(1980);
R.K.Ellis , W.J.Stirling and B.R.Webber, QCD and Collider
Physics(Cambridge University Press,1996).\\
10. W.L. van Neerven, A.Vogt, Nucl.Phys.B \textbf{588}, 345(2000).\\
11. W.L. van Neerven, A.Vogt, Nucl.Phys.B \textbf{568}, 263(2000).\\
12. S.Moch, J.Vermaseren and A.Vogt, Nucl.Phys.B \textbf{688}, 101(2004).\\
13. S.Moch, J.Vermaseren and A.Vogt, Nucl.Phys.B \textbf{691}, 129(2004).\\
14. A.Retey, J.Vermaseren , Nucl.Phys.B \textbf{604}, 281(2001).\\
15. M.M.Block, L.Durand and D.W.Mckay, Phys.Rev.D\textbf{77},
094003(2008).\\
16. E.L.Berger, M.M.Block, and Chung-I Tan,
Phys.Rev.Lett.\textbf{98},
242001(2007).\\
17. ZEUS Collaboration, V.Chekanov et.al.,
Eur.Phys.J.C\textbf{21},
443(2001).\\
18. M.Devee, B.Baishya and J.K.sarma, Eur.Phys.J.C\textbf{72},
2036(2012).\\
19. S.Weinzierl, arXiv:hep-ph/0203112.\\
20.M.M.Block, L.Durand, P.Ha and D.W.Mckay, arXiv:hep-ph/1004.1440(2010); arXiv:hep-ph/1005.2556(2010) \\
21. G.R.Boroun, JETP{\bf106}, 700(2008); B.Rezaei and G.R.Boroun,
JETP{\bf112}, 381(2011); G.R.Boroun and B.Rezaei, Eur.Phys.J.C.{\bf72}, 2221(2012).\\
22. W.L. van Neerven, E.B. Zijlstra, Phys.Lett.B{\bf272}, 127
(1991); E.B. Zijlstra, W.L. van Neerven, Phys.Lett.B{\bf273},
476(1991); Phys. Lett. B{\bf297}, 377(1992); Phys.Lett.B{\bf383}, 525(1992).\\
23. R. Hamberg, W.L. van Neerven, T. Matsuura,
Nucl.Phys.B{\bf359}, 343(1991); R. Hamberg, W.L. van Neerven, T.
Matsuura, Nucl.Phys.B{\bf644}, 403(2002), Erratum; R.V. Harlander,
W.B. Kilgore, Phys.Rev.Lett.{\bf88}, 201801(2002),
hep-ph/0201206.\\
24. B.G. Shaikhatdenov, A.V. Kotikov, V.G. Krivokhizhin, and G.
Parente, Phys.Rev.D\textbf{81}, 034008(2010).\\
25. A.Donnachie and P.V.Landshoff, Phys.Lett.B{\bf296}, 257(1992);
P.Desgrolard, M. Giffon, E.Martynov and E .Predazzi,
Eur.Phys.J.C\textbf{18}, 555(2001); P.Desgrolard, M. Giffon and
E.Martynov, Eur.Phys.J.C\textbf{7}, 655(1999); A.D.Martin,
M.G.Ryskin and G.Watt, arXiv:hep-ph/0406225.\\
26.F.D. Aaron et al. [H1 Collaboration],
Eur.Phys.J.C\textbf{71},1579(2011); C.Adloff et al. [H1
Collaboration], Eur.Phys.J.C\textbf{21}, 33(2001).\\
27.R.G.Roberts, The structure of the proton(Cambridge University Press,1990).\\
28.B.Lampe, E.Reya, Phys.Rep.\textbf{332}, 1(2000).}
\begin{figure}
\includegraphics[width=1\textwidth]{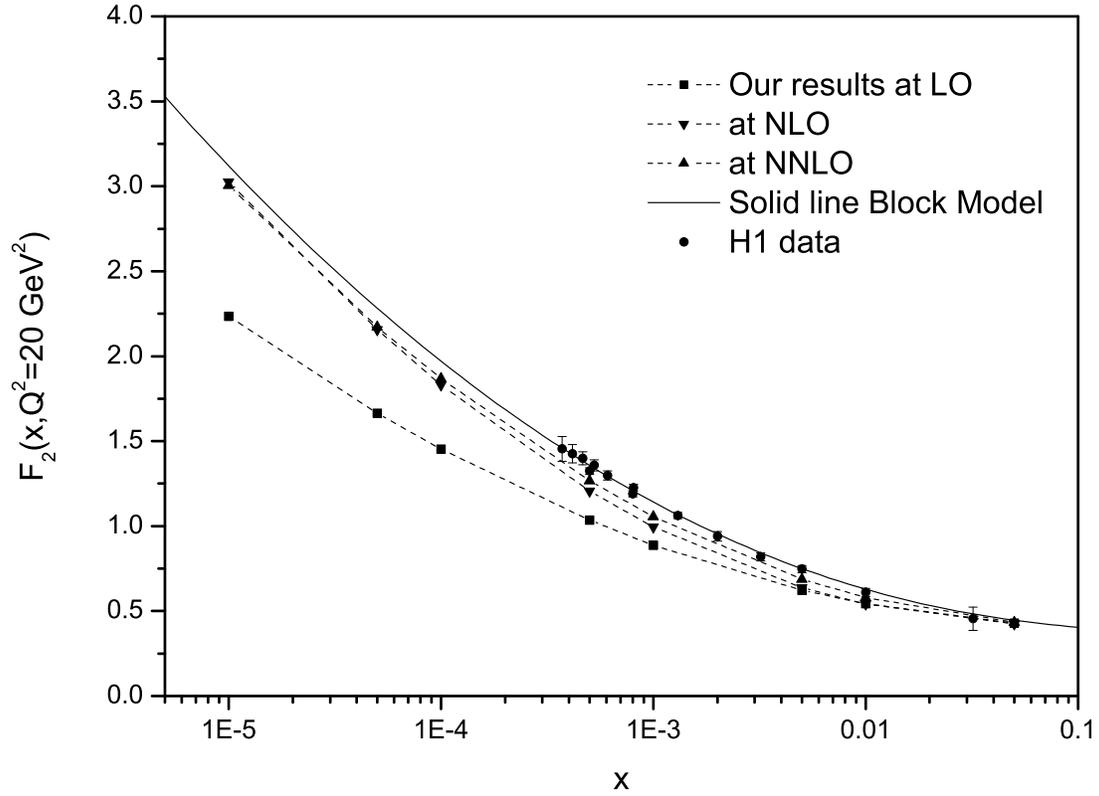}
\caption{The LO, NLO and NNLO structure functions at $Q^{2}=20
\mathrm{GeV}^{2}$. The solid curve is from Block fit [15-16]. The
dots are H1 data that accompanied with total errors
[26].}\label{Fig1}
\end{figure}
\begin{figure}
\includegraphics[width=1\textwidth]{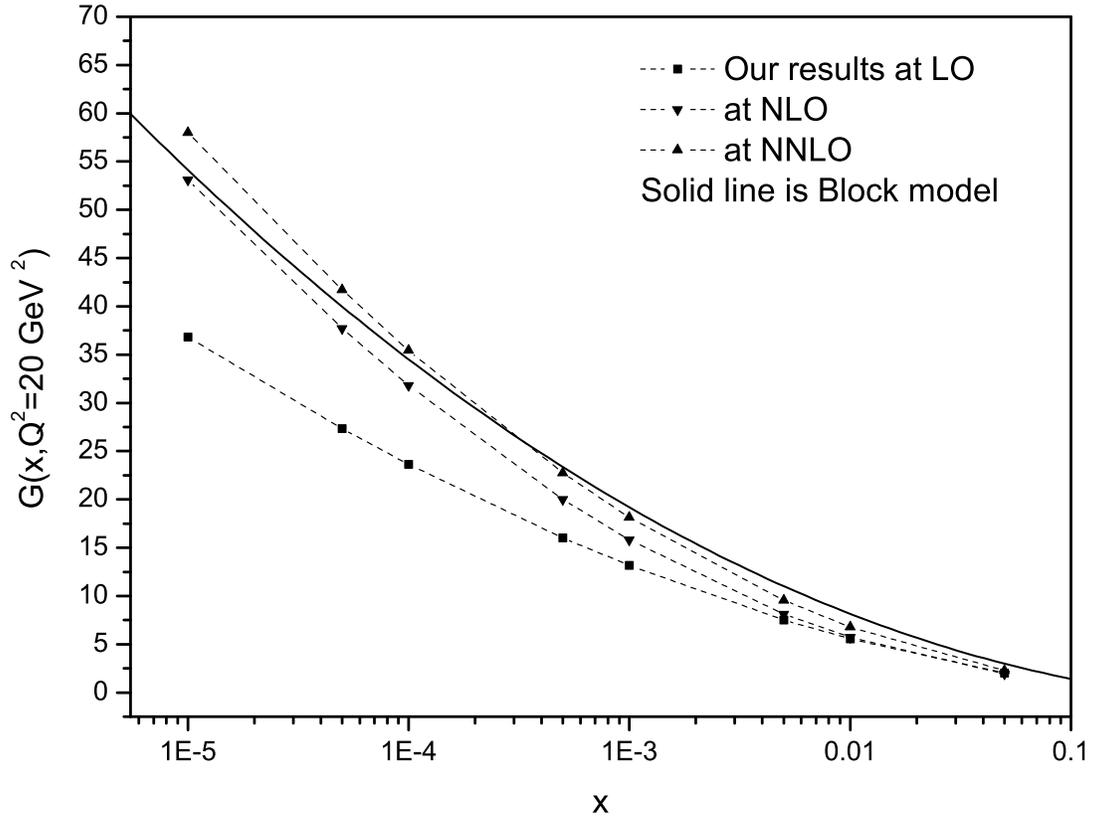}
\caption{The LO, NLO and NNLO gluon distribution function at
$Q^{2}=20 \mathrm{GeV}^{2}$. The solid curve is from Block fit
[15-16]. }\label{Fig2}
\end{figure}

\end{document}